\documentclass[12pt,preprint]{aastex}
\usepackage{url}

\begin{document}

\title{The Powerful Jet and Gamma-Ray Flare of the Quasar PKS 0438$-$436}
\author{Brian Punsly\altaffilmark{1}, Andrea Tramacere\altaffilmark{2}, Preeti Kharb\altaffilmark{3} and Paola Marziani\altaffilmark{4}}
\altaffiltext{1}{1415 Granvia Altamira, Palos Verdes Estates CA, USA
90274: ICRANet, Piazza della Repubblica 10 Pescara 65100, Italy and
ICRA, Physics Department, University La Sapienza, Roma, Italy,
brian.punsly@cox.net}\altaffiltext{2}{Department of Astronomy, University of Geneva, Chemin d'Ecogia 16 - 1290 - Versoix  - Switzerland}\altaffiltext{3}{National Centre for Radio Astrophysics,
Tata Institute of Fundamental Research, Post Bag 3, Ganeshkhind,
Pune 411007, India}\altaffiltext{4}{INAF, Osservatorio Astronomico
di Padova, Italia}

\begin{abstract}
PKS 0438$-$436 at a redshift of $z=2.856$ has been previously
recognized as possessing perhaps the most luminous known synchrotron
jet. Little is known about this source since the maximum elevation
above the horizon is low for the Very Large Array (VLA). We present
the first VLA radio image that detects the radio lobes. We use both
the 151 MHz luminosity, as a surrogate for the isotropic radio lobe
luminosity, and the lobe flux density from the radio image to
estimate a long term, time averaged, jet power, $\overline{Q}
=1.5\pm 0.7 \times 10^{47} \rm{ergs~s^{-1}}$. We analyze two deep
optical spectra with strong broad emission lines and estimate the
thermal bolometric luminosity of the accretion flow, $L_{\rm{bol}} =
6.7 \pm 3.0 \times 10^{46} \rm{ergs~s^{-1}}$. The ratio,
$\overline{Q}/L_{\rm{bol}} = 3.3 \pm 2.6 $, is at the limit of this
empirical metric of jet dominance seen in radio loud quasars and
this is the most luminous accretion flow to have this limiting
behavior. Despite being a very luminous blazar, it previously had no
$\gamma$-ray detections (EGRET, AGILE or FERMI) until December 11 -
13 2016 (54 hours) when FERMI detected a flare that we analyze here.
The isotropic apparent luminosity from 100 MeV - 100 GeV rivals the
most luminous detected blazar flares (averaged over 18 hours), $\sim
5-6 \times 10^{49} \rm{ergs~s^{-1}}$. The $\gamma$-ray luminosity
varies over time by two orders of magnitude, highlighting the
extreme role of Doppler abberation and geometric alignment in
producing the inverse Compton emission.
\end{abstract}

\keywords{black hole physics --- galaxies: jets---galaxies: active
--- accretion, accretion disks}

\section{Introduction} The quasar PKS~0438$-$436 at z = 2.856 has been
identified as the radio source with the highest radio luminosity and
the second strongest known synchrotron core \citep{mor78,pun95}.
PKS~0438$-$436 is considered a blazar due to a flat spectrum
(defined in terms of the flux density as $F_{\nu} \sim
\nu^{-\alpha}$) with $\alpha <0.5$ and its optical polarization
\citep{tin03,imp90}. Not only are there epochs of high polarization,
but it is highly variable, changing from 4.7\% to 1.7\% with a 150
degree swing in position angle in 7 months \citep{imp90,fug88}.
However, what is unusual is the enormous low frequency flux density
compared to the most luminous blazars \citep{pun95}. The 160 MHz
flux density is 7.9~Jy at z = 2.856 \citep{sle95}. For comparison,
we compute the 151 MHz luminosity $L_{151}\approx 4.5 \times 10^{28}
W \rm{Hz}^{-1}\rm{sr}^{-1}$. This is the very rare object that lies
near the high end of the luminosity distribution at both low
frequency and high frequency ($\geq 5$ GHz)
\citep{mor78,pun95,wil99}.

\par There are two unusual aspects of this luminous blazar, the large $L_{151}$ and the
lack of a detection in the gamma rays, until now. A large $L_{151}$
is typically indicative of luminous radio lobes in which energy is
stored \citep{raw91}. This study has three major objectives. First,
determine the radio lobe morphology and flux density so we can
estimate the long term time averaged radio jet power, $\overline{Q}$
(Section 2). Then, analyze the optical spectrum to get an estimate
of the thermal bolometric luminosity of the accretion flow,
$L_{\rm{bol}}$ (Section 3). In Section 4, we analyze the December
2016 gamma ray flare. In this paper we adopt the following
cosmological parameters: $H_{0}$=70 km~s$^{-1}$~Mpc$^{-1}$,
$\Omega_{\Lambda}=0.7$ and $\Omega_{m}=0.3$.

\begin{figure}
\includegraphics[width=80 mm, angle= 0]{f1.ps}
\hspace{-0.4cm}
\includegraphics[width=115 mm, angle= 0]{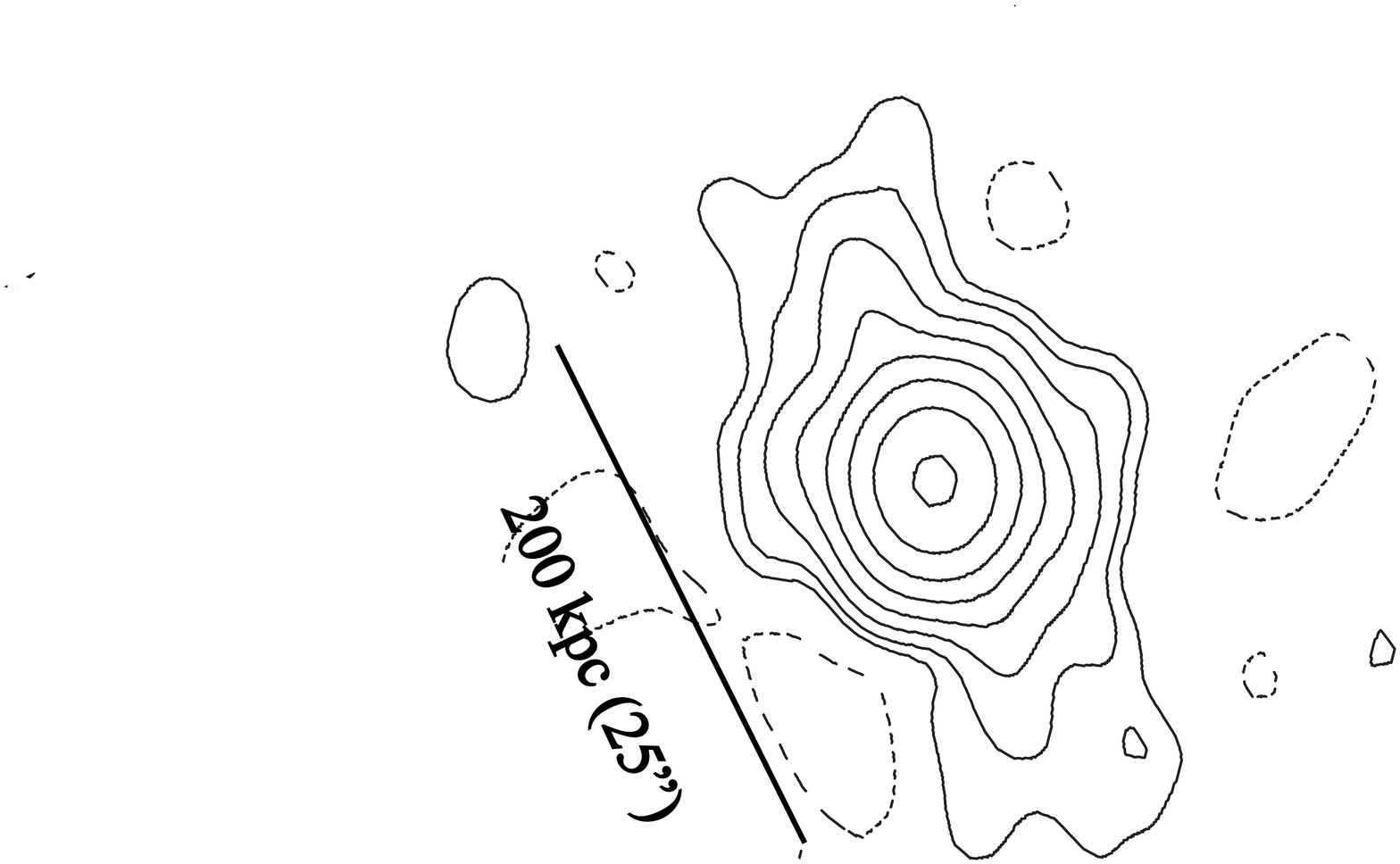}
\caption{The left panel is the JVLA image at 2.5 GHz. The image is
difficult to interpret, initially, due to the elongated beam shape
and the compact east-west morphology. The right panel is a
qualitative depiction of the intrinsic symmetry of the radio source
that is designed to clarify the intrinsic morphology of the left
hand image. It is an attempt to correct for a line of sight near the
jet axis and a highly elongated beam. It assumes that the line of
sight to the symmetry axis is $\psi=15^{\circ}$, which is used to
simulate a rotation of the radio source into the sky plane. More
details are provided in the text on the method and somewhat
arbitrary choice for $\psi$. The radio core has not been Doppler
de-boosted, so it is much more prominent than it would actually
appear in the sky plane. In spite of the limitations, the right hand
panel clarifies the intrinsic morphology of the radio source. Two
diffuse structures exist to the north-east and south-west of the
core, the radio lobes. This depiction helps one to visualize the
lobes in the image on the left. We conclude that the source in the
image in the left hand panel is likely a classical double lobe radio
source that is being viewed near the jet axis.}
\end{figure}

\section{The Large Scale Radio Structure and the Jet Power}
Estimates of instantaneous blazar jet power are often based on radio
core and $\gamma$-ray emission (eg. the methods of \citet{ghi10})
and the estimated Doppler enhancement factor, $\mathcal{D}$, can
introduce very significant uncertainty \citep{lin85,pun05}.
Estimating $\overline{Q}$ from the lobe luminosity is not
instantaneous power. Yet, it has the virtue that the lobe luminosity
and therefore $\overline{Q}$ is not sensitive to the large
uncertainty of $\mathcal{D}$ \citep{raw91,wil99}.

\subsection{$\overline{Q}$ Estimates from Archival Low Frequency Flux Density}
The estimate of $\overline{Q}$ in this paper relies on the methods
of \citet{raw91,wil99,blu00}, and will be referred to as the Oxford
Method, hereafter. The estimator for $\overline{Q}$ was calibrated
with a sample of 170 double lobe radio sources that were selected on
the basis of low frequency emission (151 MHz - 178 MHz). The
calibrated estimator depends on only one parameter, a single value
of flux density that is used as a surrogate for the flux density
restricted to the radio lobes. This is the strength of the method, a
single dish measurement at low frequency can be used to estimate
$\overline{Q}$ for sources not in the calibration sample, as long as
one can show that the measurement represents the lobe flux density.
Thus, a single dish measurement is not applicable to blazars in
general due to dilution of the lobe flux density with core and jet
flux density.

The choice of $L_{151}$ as a surrogate for the luminosity of the
radio lobes is motivated by the assumption that the core emission is
attenuated by synchrotron self-absorption at 151 MHz \citep{wil99}.
We note that the sources in the calibration sample excluded known
blazars since this would skew the calibration because there can be
significant core and jet emission at 151 MHz for a blazar. The
energy stored in the lobes, $U$, can be estimated from $L_{151}$ and
since the lobe plasma is low velocity this value will not be
strongly affected by $\mathcal{D}$ \citep{wil99}. This is the sole
function of the parameter $L_{151}$ in the Oxford Method. Consider
the energy conservation equation (this is Equation (4) of
\citet{wil99}, EQW4, hereafter),
\begin{equation}
\overline{Q} = \frac{fU_{\rm{base}}(L_{151})}{T}\;,
\end{equation}
where $T$ is the source age, $U_{\rm{base}}(L_{151})$ is the minimum
energy in the lobe assuming a low frequency cut off at 10 MHz, the
jet axis is $60^{\circ}$ to the line of sight, there is no protonic
contribution and 100\% filling factor. \footnote{Since we will
ultimately be applying these formulae to blazars, the choice of the
jet axis being $60^{\circ}$ to the line of sight needs some
explanation. This is only used with regard to the calibration sample
of double lobe radio sources. It is used to estimate the volume in
the lobes It has nothing to do with the line of sight for which the
formula can be applied. For example, double lobe quasars are
believed to have a line of sight to the jet axis of $< 45^{\circ}$
in the unified scheme for radio sources \citep{bar89}. Yet, the
formula is routinely applied to quasars in the literature as it was
intended.}The quantity $f$ incorporates deviations of actual radio
lobes from these assumptions as well as energy lost expanding the
lobe into the external medium, back flow from the head of the lobe
and kinetic turbulence. Secondly, Equation (9) of \citet{wil99}
relating $T$ and $\overline{Q}$ (EQW9, hereafter) is derived by
considering the evolution of the lobe dimension, $R$, in models of
lobe head advance into an ambient medium. Differentiating EQW9
yields an equation for $dR/dt$ (EQW10, hereafter) that is equated to
the lobe advance speeds derived from the distribution of length
asymmetry between the two lobes of a sample of quasars in
\citet{sch95}. EQW4, EQW9, EQW10 were solved simultaneously to
eliminate $R$ and $T$, yielding $\overline{Q}$ as a function of $f$
and $L_{151}$ that is plotted in a scatter plot of the calibration
sample in Figure 7 of \citet{wil99},
\begin{equation}
\overline{Q} \approx 3.8\times10^{45} f L_{151}^{6/7} \rm{ergs/s}\;,
\end{equation}
The exponent on $f$ is 1 not 3/2 as in \citet{mar11}
\par A more detailed effort to constrain
$f$ that included gradients in the lobe magnetic field and
significant deviations from equipartition estimated that $10<f<20$
\citep{blu00}. They assume a ``relaxed" classical double lobe radio
source in the derivation of their equations. Implicit in the relaxed
constraint that was implemented in the determination of $f$ is that
the linear size of the radio source is larger than the size of the
host galaxy (i.e. $> 20$ kpc). Thus, the lobes are not confined by
the pressure of the relatively dense galactic medium. The assumed
configuration is also implemented because a classical double lobe
radio source is an object with a jet that has significant
inclination to the line of sight (therefore small $\mathcal{D}$)
that is dominated by two luminous lobes with a faint jet and core.
In this configuration, $L_{151}$ is indeed an excellent surrogate
for the lobe luminosity. We bound $\overline{Q}$ in Equations (2)
from above (below) by using $f=20$ ($f=10$). Using $L_{151}\approx
4.5 \times 10^{28} W \rm{Hz}^{-1}\rm{sr}^{-1}$ from Section~1 in
Equation (1),
\begin{equation}
\overline{Q} = 1.65\pm 0.55\times 10^{47} \rm{erg/s} \;.
\end{equation}
\subsection{Applying the Oxford Method to Blazars}
In \citet{pun05}, it was shown that one does not need a single dish
measurement to estimate the lobe luminosity. A low frequency, high
resolution, high sensitivity radio image can actually be used to
determine the flux density in the lobes directly and this is
equivalent to a single dish measurement in lobe dominated sources.
More importantly, a high dynamic range image can be used to isolate
the radio lobes from a blazar like radio core and jet, allowing a
robust estimate of the flux density of the radio lobes. The
additional data supplied by the radio image allows the Oxford Method
to be applied to blazars. Recall that the sole purpose of $L_{151}$
in the derivation of Equation (2) was to estimate
$U_{\rm{base}}(L_{151})$ in Equation (1). According to Equation (1)
in \citet{wil99}, the defining relation for
$U_{\rm{base}}(L_{151})$, the minimum energy only depends on the
combination $F_{\nu}\nu^{\alpha}$ which is a constant for a power
law. Thus, it makes no difference if the one uses 2.5 GHz or 151 MHz
to estimate $U$. Both choices have an uncertainty associated with
$\alpha$ that is captured in the calibration of $f$.

\par Unfortunately, due to the low declination, the maximum elevation above the horizon is low
in New Mexico and PKS 0438$-$436 is visible for very short durations
with the Very Large Array (VLA). The old VLA had insufficient u-v
coverage to image the diffuse lobes during these brief observations.
However, with the large bandwidths of the new Jansky VLA (JVLA), the
u-v coverage is much better and the sensitivity much higher. PKS
0438$-$436 was observed for 5-6 minutes throughout the summer of
2015 as part of Project 15A-105 as a phase calibrator. We utilize
these observations in order to image the kpc structure of PKS
0438$-$436 for the first time and improve the $\overline{Q}$
estimate. \subsubsection{Absolute Flux Calibration} We used PKS
0438$-$436 not only as its own phase calibrator, but its own
amplitude calibrator. In general this scheme seems flawed due to
blazar variability. However, PKS 0438$-$436 is not that variable at
lower frequency (eg., 2.5 GHz). Our frequency of observation is 2.5
GHz in A-array. We found 18 ATCA (Australian Telescope Compact
Array) observations from 1997 - 2011, from the ATCA calibrator
web-page\footnote{http://www.narrabri.atnf.csiro.au/calibrators/}
and \citet{tin03,mas08}, all indicating a narrow range of flux
density, $3-4$~Jy.
\par We combine this steady nature of the flux with a fortuitous ATCA
calibrator observation from 4.4~GHz to 6.4~GHz on August 13 2015.
The extrapolated 2.5~GHz flux density was $3.25\pm0.33$~Jy. There
are JVLA observations on August 15 and August 24, 2015. The August
24 observation is imaged in Figure 1 since it had more uniform and
symmetric coverage in the {\tt uv} plane and this is the most
important consideration for detecting diffuse, extended emission.
\par Those data were reduced using the EVLA pipeline incorporated in
CASA version 4.7.2. As the data did not include a flux or bandpass
calibrator, the pipeline turned out to be useful mostly for the
editing (flagging) of bad data, which it successfully completed for
all the 96 spectral windows ({\tt spw}) and 64 channels. The
corrected  data were separated from the main dataset using the task
{\tt SPLIT} and imaged using {\tt CLEAN} in the multi-frequency
synthesis ({\tt mfs}) widefield mode, with the number of Taylor
coefficients ({\tt nterms}) being 2 and number of w-projection
planes being 128; ``Briggs'' weighting scheme was used for the
imaging. One round of phase-only self-calibration was carried out
before creating the final image.
Finally, the AIPS verb {\tt RESCALE} was used to correct the flux
density scale. The core parameters using the AIPS Gaussian-fitting
task JMFIT were: peak intensity = 2.74~Jy~beam$^{-1}$ and the
integral intensity = 2.82~Jy (see the left hand panel of Figure1).
The extended flux density is $0.432\pm0.043$ Jy, where the error is
from the uncertainty in the amplitude calibrator.

\subsubsection{The Nature of the Radio Lobes and Jet Power}
We note that north-east and south-west extensions in emission were
detected in 327 MHz Very Large Baseline Array (VLBA, hereafter)
observations of \citet{kan09}. These features appear to be a
partially resolved jet and counter-jet on scales $\sim 100$ mas from
the unresolved radio core. These smaller scale features are buried
deep within the unresolved JVLA radio core (i.e., it is an order of
magnitude smaller than the resolution of the 2.5 GHz observations).
The jet and counter-jet directions detected with the VLBA are
consistent with jets that connect to the extended emission in Figure
1. This supports the notion that the partially resolved structures
in the left hand panel of Figure 1 are distant radio lobes.

\par In Section 2.1, we discussed how the Oxford Method for estimating $\overline{Q}$
in Equation (2) is derived under the assumption of a relaxed
classical double radio source. Thus motivated, we aim to establish
that PKS~0438$-$436 is a classical double radio source except for
the near polar line of sight. The elongated beam shape due to the
proximity to the horizon makes the pole-on geometry difficult to
interpret. The apparent separation of the radio lobes is about 50
kpc. However, this is not likely indicative of the actual physical
separation of the lobes due to projection effects. Blazars that are
high polarization quasars (HPQs), such as PKS~0438$-$436, have lines
of sight (LOS) to the base of the jet that have been estimated from
time variability brightness temperatures and the ``inverse Compton
catastrophe" to be $\approx 3.3^{\circ}$ \citep{hov09}. However, the
jet is highly curved based on multi-frequency VLBA observations. The
position angle of the jet direction rotates $\approx 90^{\circ}$
between 35 mas and 110 mas from the core \citep{kan09,tin98,pus12}.
Thus, one does not necessarily expect the same angle between the LOS
and the jet at 110 mas and 2 arcsec from the core as the angle of
the LOS to the jet base ($\ll 100$ mas from the point of origin).
The 327 MHz VLBA detection of a counter-jet to the south-west is
inconsistent with the $\mathcal{D}$ associated with blazar activity.
A possible explanation is the curving jet that results in a LOS,
$\psi$, that is less aligned with the jet on scales $\sim 100$ mas
and larger than that of an extreme blazar ($\approx 3.3^{\circ}$),
thereby reducing $\mathcal{D}$ \citep{lin85}. According to the
unification scheme of FR II (Fanaroff Riley II) radio sources, the
average line of sight to the $>10$ kpc jet in a quasar is $\psi
\approx 30^{\circ}$ \citep{bar89}. The curved jet and counter-jet
detection suggest a LOS larger than the HPQ value, $\psi >
3.3^{\circ}$, but we still assume that it is more blazar-like than a
typical radio loud quasar. Thus, we pick an intermediate value of
$\psi \approx 15^{\circ}$ as a plausible value in order to
qualitatively explore the intrinsic morphology of the radio source.

\par Even if one assumes no interaction with the enveloping medium,
comparing parameters of extended features such as lobes on the jet
and counter-jet side of the core (i.e., flux or displacement from
the core), one can determine at most the combination,
$\beta_{\rm{lobe}}\cos{\psi}$, for an individual object, where
$\beta_{\rm{lobe}}$ is the lobe advance speed (normalized to the
speed of light) and $\psi$ is the angle of the line of sight to the
lobe propagation direction \citep{sch95,blu00,lin85}. The situation
is exacerbated by a source without high signal to noise images of
the radio lobes. First consider, the ratio of lobe separation on the
jet (approaching) and counter-jet (receding) sides of the core
$L_{\rm{asym}}$. We approximate $L_{\rm{asym}} \approx 1.05$ based
on the radio image. Theoretically assuming intrinsic bilateral
symmetry and no interaction with the enveloping medium,
$L_{\rm{asym}}
=(1+\beta_{\rm{lobe}}\cos{\psi})/(1-\beta_{\rm{lobe}}\cos{\psi})$
\citep{sch95}. The empirical estimated asymmetry of 1.05 and $\psi
\approx 15^{\circ}$ implies that $\beta_{\rm{lobe}}\approx 0.03$.
Similarly, the ratio of flux density of the jetted lobe to
counter-jetted lobe is estimated from the radio image as
$F_{\rm{asym}}\approx 1.8$. Theoretically, for a spherical region,
intrinsic bilateral symmetry and minimal interaction with the
enveloping medium,
$\beta_{\rm{lobe}}\cos{\psi}=[(F_{\rm{asym}})^{1/(3+\alpha)}+1]/[(F_{\rm{asym}})^{1/(3+\alpha)}-1]$
\citep{rib04}. For a steep lobe spectral index of $\alpha=0.9$
\citep{kel69}, one finds for $F_{\rm{asym}}\approx 1.8$ and $\psi
\approx 15^{\circ}$ that $\beta_{\rm{lobe}}\approx 0.08$. The values
of $\beta_{\rm{lobe}}=0.03$ and $\beta_{\rm{lobe}}=0.08$, from
$L_{\rm{asym}}$ and $F_{\rm{asym}}$, respectively agree with the
findings of \citet{sch95} of a most common value of
$\beta_{\rm{lobe}}= 0.03 \pm 0.02$ and $\beta_{\rm{lobe}}<0.15$ in
FR II radio sources. This argument merely shows that $\psi \approx
15^{\circ}$ is consistent with what is known about lobe advance
speeds, but it does not prove that this is a unique value. $\psi$
cannot be determined from the existing data. If $\psi$ were
$\psi=10^{\circ}$ or $\psi = 30^{\circ}$, the intrinsic length would
be $\approx 350$ kpc or $\approx 100$ kpc, respectively; yet the
fundamental conclusion that the source is inherently a classical
double lobe radio source being viewed along a LOS nearly parallel
with the axis of the jet base is unchanged.

\par For illustrative purposes, we stretch the image so it appears as
if the radio lobes have been rotated into the sky plane. This
exercise is designed to elucidate the likelihood that PKS~0438$-$436
is inherently a classical double lobe radio source that happens to
be viewed nearly pole-on to the base of the jet and therefore
appears as a blazar within the standard unification scheme of FR II
radio sources \citep{bar89}. The method of altering the image of
PKS~0438$-$436, so that the lobes appear as they would if they were
to lie in the sky plane instead of at $\psi \approx 15^{\circ}$ is
fairly simple. The image in the left hand panel of Figure 1 is
stretched by a factor of $1/\sin{15^{\circ}} \approx 4$ along the
symmetry axis (the imaginary lines connecting the two lobes and the
core) in Adobe Illustrator. The result is displayed in the right
hand panel of Figure 1. This crude simulation does not correct for
$\mathcal{D}$, so the core is much brighter than it would be if we
were observing PKS 0438$-$436 in the sky plane. The implication of
the figure (in spite of its limitations) is that PKS 0438$-$436
would be a classical double lobe radio sources if not for the fact
that it is being viewed almost pole-on and it is being imaged with
the added distortion of a very asymmetric beam. The choice of $\psi
\approx 15^{\circ}$ does not affect this conclusion, only the
intrinsic size of the object is affected by this choice. The
stretched image highlights the existence of two lobes of extended
emission that are approximately equidistant from the core and well
outside of the galactic dimensions. These are the characteristics of
a relaxed classical double radio source. Thus, the luminosity of
these lobes are precisely the input that is intended to be used in
the Oxford Method formula, Equation (2). The fact that we have not
determined the exact intrinsic linear size or the actual value of
$\psi$ is irrelevant to the intended use of the formula as discussed
at length in Section 2.1
\par We assume that for the radio lobes $\alpha_{\rm{lobe}} = 0.9$ \citep{kel69}. Combining
this with our estimate of the lobe flux density at 2.5 GHz above and
Equation (2),
\begin{equation}
L_{151}\approx 3.0 \times 10^{28} W \rm{Hz}^{-1}\rm{sr}^{-1}\;,
\quad \overline{Q} = 1.14 \pm 0.38 \times 10^{47} \rm{erg/s} \;.
\end{equation}
This seems to indicate that Equation (3) is an overestimate due to
core and jet contributions to the 160 MHz flux density. Equation (4)
might be more accurate since the method of \citet{wil99} assumes
that $L_{151}$ is a surrogate for the lobe luminosity. If
$\alpha_{\rm{lobe}}$ were 1.05 then the 7.9 Jy single dish flux
density at 160 MHz could be attributed entirely to the lobes.
However, this steep of a spectrum is an extremely rare circumstance
\citep{kel69}.

\begin{figure}
\begin{center}
\includegraphics[width=135 mm, angle= 0]{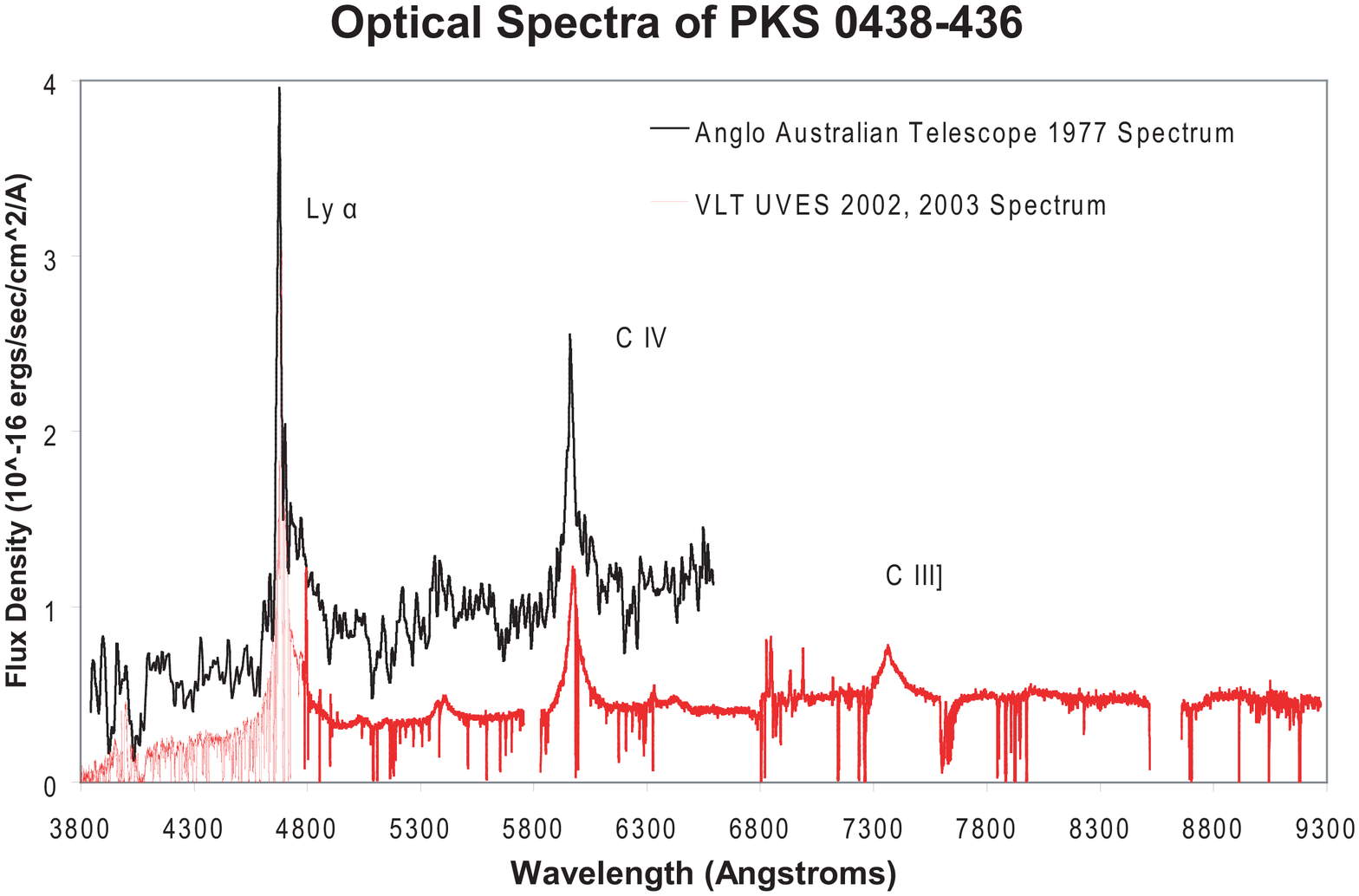}
\includegraphics[width=75 mm, angle= 0]{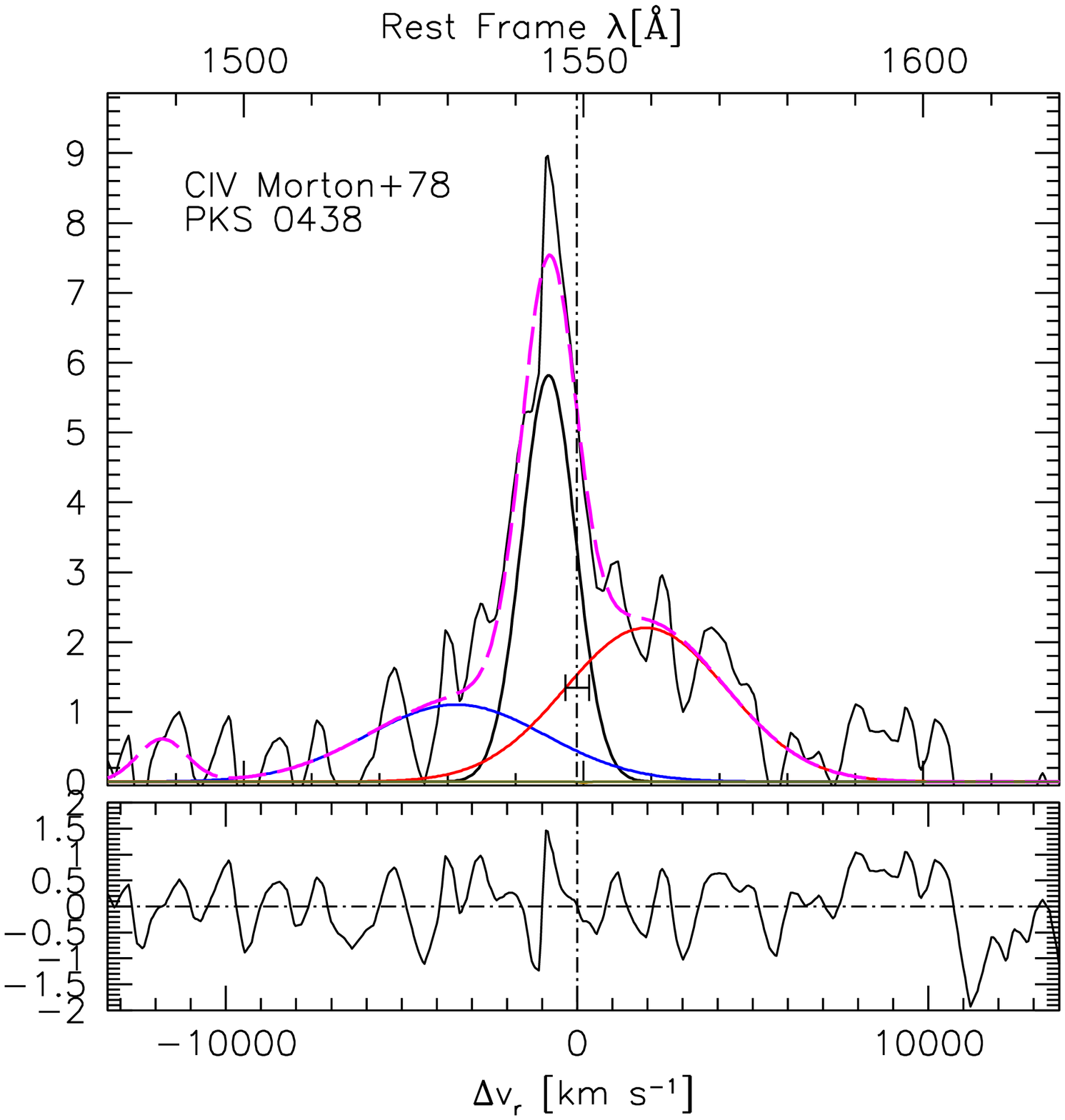}
\includegraphics[width=75 mm, angle= 0]{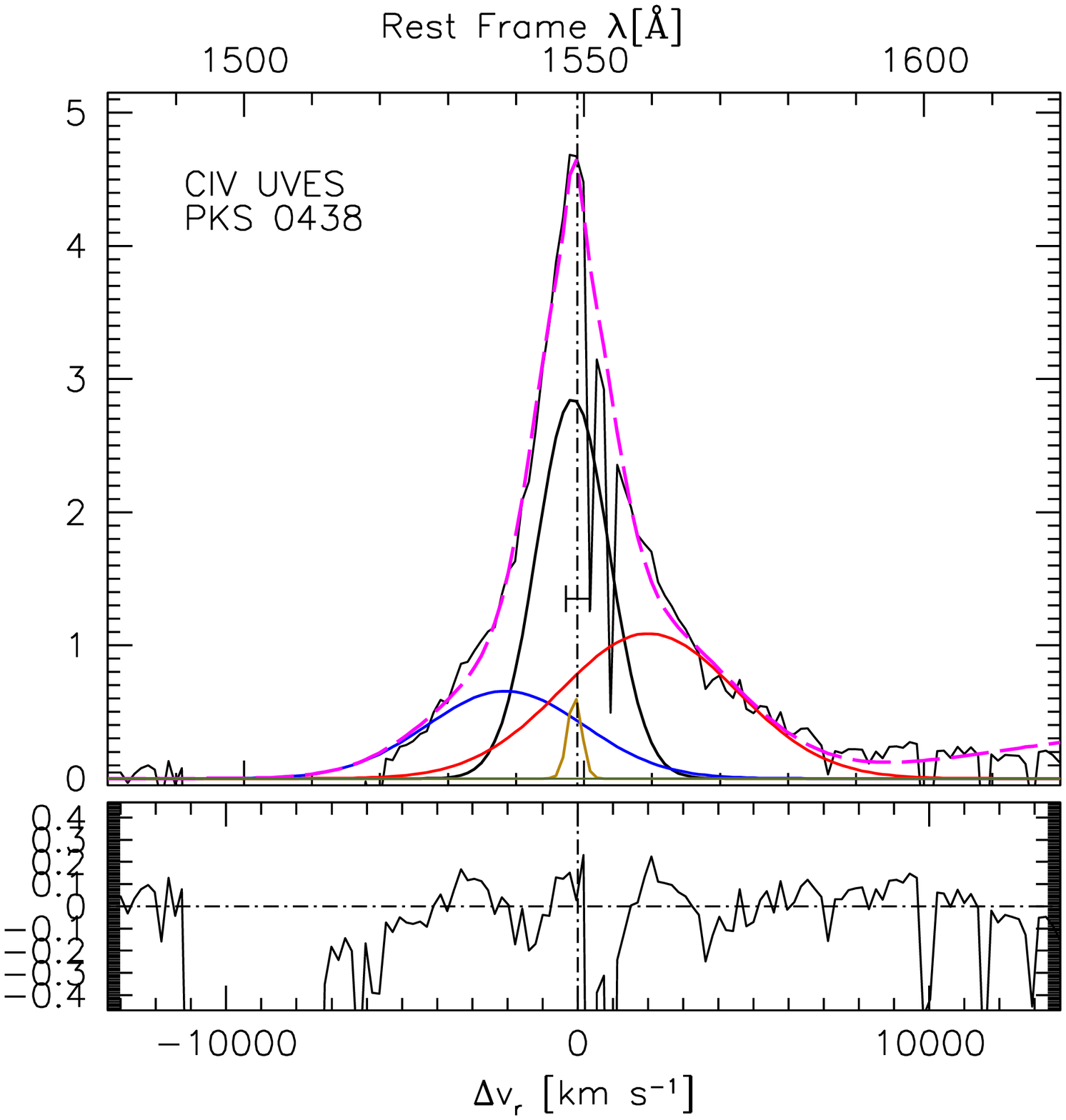}
\caption{The top frame shows the two spectra. The bottom left panel
is the CIV fit from the 1977 observation and the bottom right panel
is the fit to the 2003 CIV.}
\end{center}
\end{figure}
\section{The Luminosity of the Accretion Flow} In this section, we examine optical spectra in order to
quantify the thermal luminosity of the accretion flow and the
strength of the broad emission lines. Due to the high redshift, the
UV emission line spectrum is redshifted into the optical band. The
earlier spectrum (November 18 1977) in Figure 2 is from the RGO
spectrograph on the 3.9 m Anglo Australian Telescope and this
observation was used in the original determination of the quasar
properties and redshift \citep{mor78}. The spectra from UVES, the
Ultraviolet and Visual Echelle Spectrograph, on the VLT were
retrieved from the ESO Spectral Data Products of Phase 3 archive
portal. The UVES spectrum of Figure 2 results from the combination
of eleven flux-calibrated spectra. On Sept. 14, 2002 a total of
eight spectra were taken: four covering $3730\AA$ -$5000\AA$ and
four covering $6650\AA$ -$10420\AA$. On Oct. 30, 2003 three spectra
covering $4720\AA$ -$6830\AA$ were combined. A gap is present on the
blue side of the redshifted CIV$\lambda$1549, but the profile is
presumably unaffected. There is also a small absolute flux mismatch
around $6800\AA$ between 2002 and 2003. The spectral resolution of
UVES is about R = $\lambda$/$\delta\lambda \sim$ 40,000 with the
slit opened to 1.008". We re-binned the spectra to a uniform scale
0.04 \AA/pixel in advance of the multi-component minimum $\chi^2$
best fitting analysis.

The luminosity appears larger in 1977 the black plot in the top
frame of Figure 2). However, it should be noted that a significant
flux correction was applied in \citet{mor78} to compensate for $\sim
3.5"$ seeing and a 1.25" aperture. The rising spectrum with
wavelength and higher flux level indicates that a strong optical
synchrotron jet dominates the continuum. The spectral index is
$\alpha_{\nu} \approx 3.4$, $F_{\nu}\sim \nu^{-\alpha_{\nu}}$
\citep{mor78}. Note that the plot in Figure 2 is with respect to
wavelength. We confirm the Morton calculation. We estimate the flux
density of the continuum as $F_{\lambda}(\lambda=6600 \AA) \approx
1.2\times 10^{-16} \rm{ergs/sec}/\rm{cm}^{2}/\AA$ and
$F_{\lambda}(\lambda=4100 \AA) \approx 6.0\times 10^{-17}
\rm{ergs/sec}/\rm{cm}^{2}/\AA$. Define the spectral index with
respect to wavelength, $\alpha_{\lambda}$, as $F_{\lambda} \propto
\lambda^{\alpha_{\lambda}}$. Then $\alpha_{\lambda}\approx
\log(1.2/0.6)/\log(6600/4100) \approx 1.45$. $\alpha_{\nu} = 2 +
\alpha_{\lambda} \approx 3.45$ in close agreement with the published
result \citep{mor78}.
\par The emission line fits are presented in Table 1 and the CIV fits are
shown in the bottom panels of Figure 2. The emission line fit was
done simultaneously with the continuum power law fit. We used a
$\chi^{2}$ minimization technique in IRAF with the task ``specfit."

\begin{table}
\caption{Broad Emission Line Fits}
{\tiny\begin{tabular}{cccccccccccc} \tableline\rule{0mm}{3mm}
 Date &  Line & Red BC & Red BC & Red BC & Blue BC & Blue BC & Blue BC & IBL & IBL & Total BEL \\
 &   &  CF & FWHM & Luminosity & CF & FWHM &Luminosity & FWHM &Luminosity & Luminosity  \\
 &    &  km~s$^{-1}$ &  km~s$^{-1}$ & ergs~s$^{-1}$ & km~s$^{-1}$ &km~s$^{-1}$ & ergs~s$^{-1}$ & km~s$^{-1}$  & ergs~s$^{-1}$ & ergs~s$^{-1}$  \\
\tableline \rule{0mm}{3mm}
11/18/1977 & Ly $\alpha$ & $2733$ & $7143\pm 136$ &  $6.42 \times 10^{44}$& $-4558$ &$3623\pm433$  &$8.20 \times 10^{43}$ & $1311 \pm 44$& $3.91 \times 10^{44}$ & $1.11 \times 10^{45}$ \\
11/18/1977 & CIV & $2823$ & $5368\pm 538$ &  $3.11 \times 10^{44}$& $-3437$  & $6026\pm 903 $ &$1.73 \times 10^{44}$ & $1806 \pm 75$& $2.76 \times 10^{44}$ & $7.61 \times 10^{44}$ \\
9/14/2002 & Ly $\alpha$ & $1422$ & $6264\pm 491$ &  $3.16 \times 10^{44}$&$-1897$ & $6410\pm 785$  & $2.06 \times 10^{44}$ & $1747 \pm 52$& $4.06\times 10^{44}$ & $9.28\times 10^{44}$ \\
10/30/2003 & CIV & $2020$ & $5865\pm 312$ &  $1.69 \times 10^{44}$& $-2063$  & $5316\pm776$ & $9.10 \times 10^{43}$ & $2424 \pm 24$ & $1.82 \times 10^{44}$ & $4.41 \times 10^{44}$ \\
9/14/2002 & CIII] & $2164$ & $5855\pm 183$ & $8.95 \times 10^{43}$ & \tablenotemark{a} & \tablenotemark{a} & \tablenotemark{a}  & $3256 \pm 221$  & $1.01 \times 10^{44}$ & $1.91 \times 10^{44}$ \\ \hline
\end{tabular}}
\tablenotetext{a}{Not detected.}
\end{table}
We decompose the broad emission lines (BELs) into three components,
the BC \citep[also called the intermediate broad line or
IBL;][]{bro94}, the redshifted  BC \citep[very broad component
following][]{sul00} and a blueshifted BC \citep{bro96,mar96,sul00}.
The IBL/BC is represented by the black Gaussian profiles in Figure 2
and has been defined as a component with a FWHM
$\sim$2000~km~s$^{-1}$ \citep{bro94}. We occasionally add a weak
narrow line that is at the systemic velocity of the quasar (brown).
The BC needs some explanation in terms of the CIV decomposition. In
\citet{bro94}, the BC was considered to have a blueshifted center
frequency (CF) $\sim$1000~km~s$^{-1}$ and a FWHM
$\sim$7000~km~s$^{-1}$. According to Table 1, there is a component
like this in CIV, but it is less prominent than the red BC. The
blueshifted BC has been found to be dominant in radio quiet quasars,
but is  less prominent and can be completely absent in radio loud
quasars \citep{ric02,pun10}. This was later interpreted in terms of
Population B quasars in the classification scheme of \citep{sul00}.
The Population B quasars include most of the radio loud quasars and
typically have very large black hole masses and low Eddington
ratios: $R_{\rm{Edd}} \equiv L_{\rm{bol}}/L_{\rm{Edd}}\sim 0.01
-0.1$, where $L_{\rm{bol}}$ is the thermal bolometric luminosity of
the accretion flow and $L_{\rm{Edd}}= 1.26 (M_{bh}/M_{\odot})\times
10^{38}~\rm{ergs~s^{-1}}$ is the Eddington luminosity expressed in
terms of the central supermassive black hole mass, $M_{bh}$
\citep{sul07}. The Population A quasars typically have smaller FWHM
and higher Eddington ratios than Population B quasars \citep{sul07}.
This decomposition of the quasar population is consistent with
models of outwardly spiralling BEL gas that is driven by the intense
radiation pressure form the accretion flow \citep{mur95}. Population
B CIV profiles do not tend to have a prominent blue excess at their
base, but this excess in often detectable. In this respect,
PKS~0438$-$436 seems to have properties of CIV which are consistent
with the radio loud Population B quasars in terms of the
distribution of centroid line shifts \citep{sul07}. Empirically, the
blue BC might represent the existence of a moderate $R_{\rm{Edd}}$
accretion flow, and the even stronger red BC is correlated with the
extremely powerful jet \citep{pun10}.

\par We wish to estimate $L_{\rm{bol}}$ in a manner that does not include
reprocessed radiation in the infrared from distant molecular clouds.
This would be double counting the thermal accretion emission that is
reprocessed at mid-latitudes \citep{dav11}. The accretion flow
continuum may be hidden by strong synchrotron dilution. As is
commonly done for blazars, we use the BEL luminosity, $L(CIV)$, from
Table 1 as a surrogate for $L_{\rm{bol}}$ \citep{wan04}. We employ
an estimator obtained by comparing $L(CIV)$ to the Hubble Space
Telescope ultraviolet composite continuum for quasars with
$L_{\rm{bol}}\gtrsim 10^{46} \rm{ergs/sec}$,
\citep{zhe97,tel02,pun16}
\begin{eqnarray}
&&L_{\mathrm{bol}}=(107 \pm 22) L(CIV)\approx \nonumber\\
&& 4.7\pm 1.0 \times 10^{46} \rm{ergs/s}:\, \rm{October\,
2003},\quad 8.1 \pm 1.7 \times 10^{46} \rm{ergs/s}:\; \rm{November\,
1977} \;.
\end{eqnarray}
The uncertainty in Equation (5) is the uncertainty in the Hubble
Space Telescope composite continuum level and the uncertainty in
$L(CIV)$ added in quadrature \citep{tel02}. We can compare this to
an estimate from the low state continuum in the far UV in order to
see how much the steep synchrotron tail dominates in this region of
the spectrum. From the spectrum in Figure 2 and the formula
expressed in terms of quasar cosmological rest frame wavelength,
$\lambda_{e}$ and spectral luminosity, $L_{\lambda_{e}}$, from
\citet{pun16},
\begin{equation}
L_{\mathrm{bol}} \approx (4.0 \pm 0.7)
\lambda_{e}L_{\lambda_{e}}(\lambda_{e} = 1350 \AA)\approx 4.5 \pm
0.8 \times 10^{46} \rm{ergs/s} \;.
\end{equation}
This estimate from September 2002 is similar to our lower BEL
estimate in Equation (5) from 2003. Thus, just longward of
Ly$\alpha$ it appears that the accretion disk continuum is
dominating over the synchrotron component. Combining Equations (3) -
(6) we find that $\overline {Q}/L_{\mathrm{bol}}=3.3 \pm 2.6$ which
is extremely high for a quasar \citep{pun10}.

\section{The Gamma Ray Flare}
A preliminary FERMI detection of PKS 0438$-$436 on December 11 2016
was announced in ATel 9854, it previously had no gamma ray
detections; EGRET, AGILE or FERMI \citep{che16}. We have performed
the data analysis of  $Fermi$-LAT data using the fermipy framework
\citet{wood2017} version 0.16.0, based on the Fermi Science Tools
$v11r5p3$ and \verb|P8R2_SOURCE_V6| instrument response functions.
The data have been selected in the time span of six days, between
57731.0 MJD 57737.0 MJD. We have selected photons from a region of
interest (ROI) with a radius of $10 ^{\circ}$, centered on PKS
0438$-$436,  and we have built a model of the ROI using sources
within $15 ^{\circ}$ distance from the ROI center and reported in
the Fermi-LAT Third Source Catalog (3FGL) \citet{Acero2015},
including both the emission from the Galactic diffuse
(\verb|gll_iem_v06.fits|) and the isotropic
\verb|iso_P8R2_SOURCE_V6_v06.txt| backgrounds.

For the light curve extraction we have used bins of 18 hours, for a
time range starting on MJD 57731.00 and ending on MJD  57736.25. For
each  bin a binned likelihood analysis is performed, using la
power-law function, and freeing all the sources located within a
distance of $5 ^{\circ}$ from the ROI center. The light curve is
shown in  Figure 3, top frame. The top left frame reports the
integrated photon flux in the 100 MeV - 100 GeV energy range, and
the top right frame reports the corresponding apparent luminosity.
The source is detected in four bins with a significance $>3 \sigma$,
and the flaring activity in concentrated during a time span of 54
hours centered on
 MJD 57734.4.

\begin{figure}
\begin{center}
\includegraphics[width=100 mm, angle= 0]{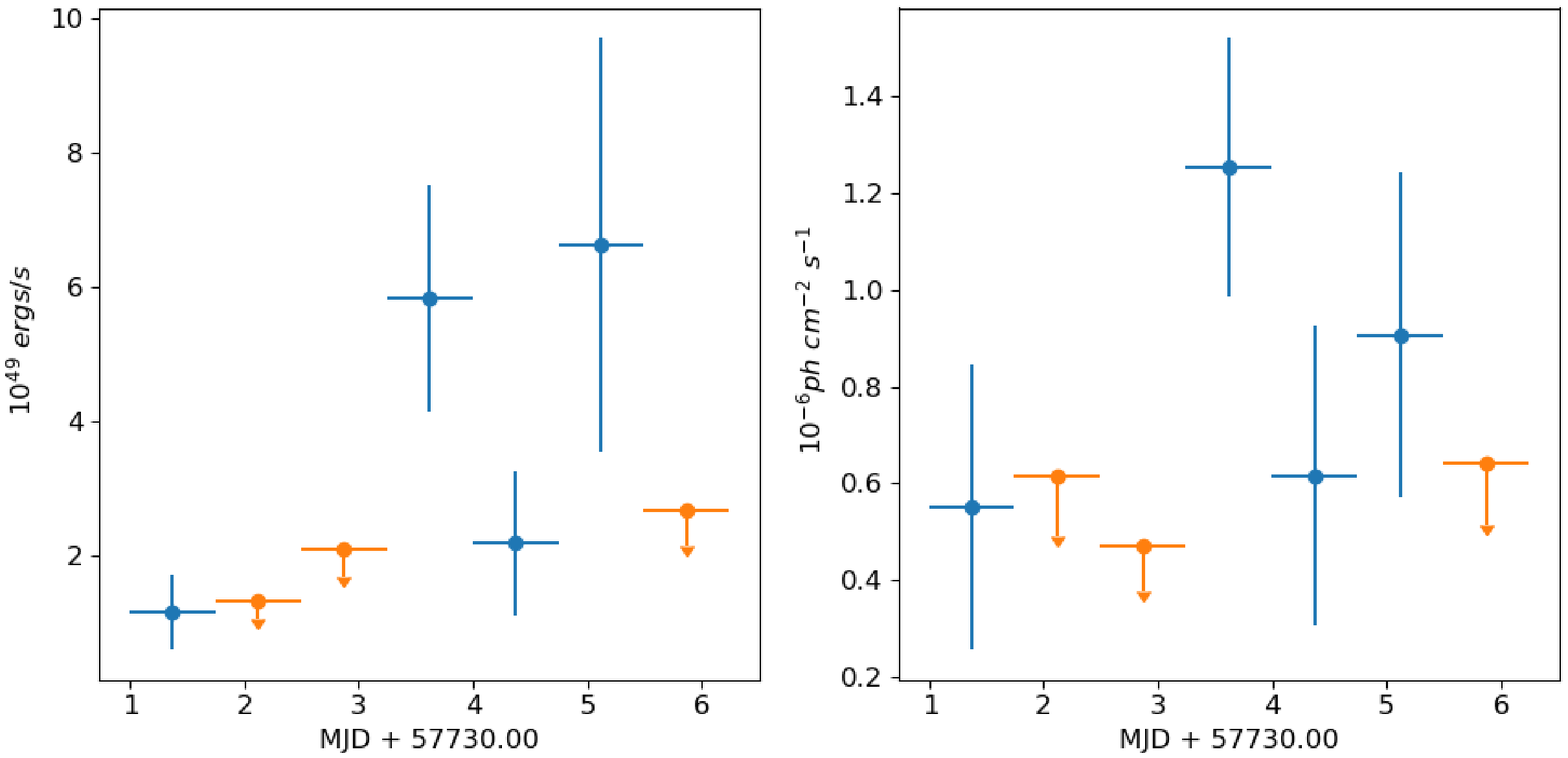}
\includegraphics[width=65 mm, angle= 0]{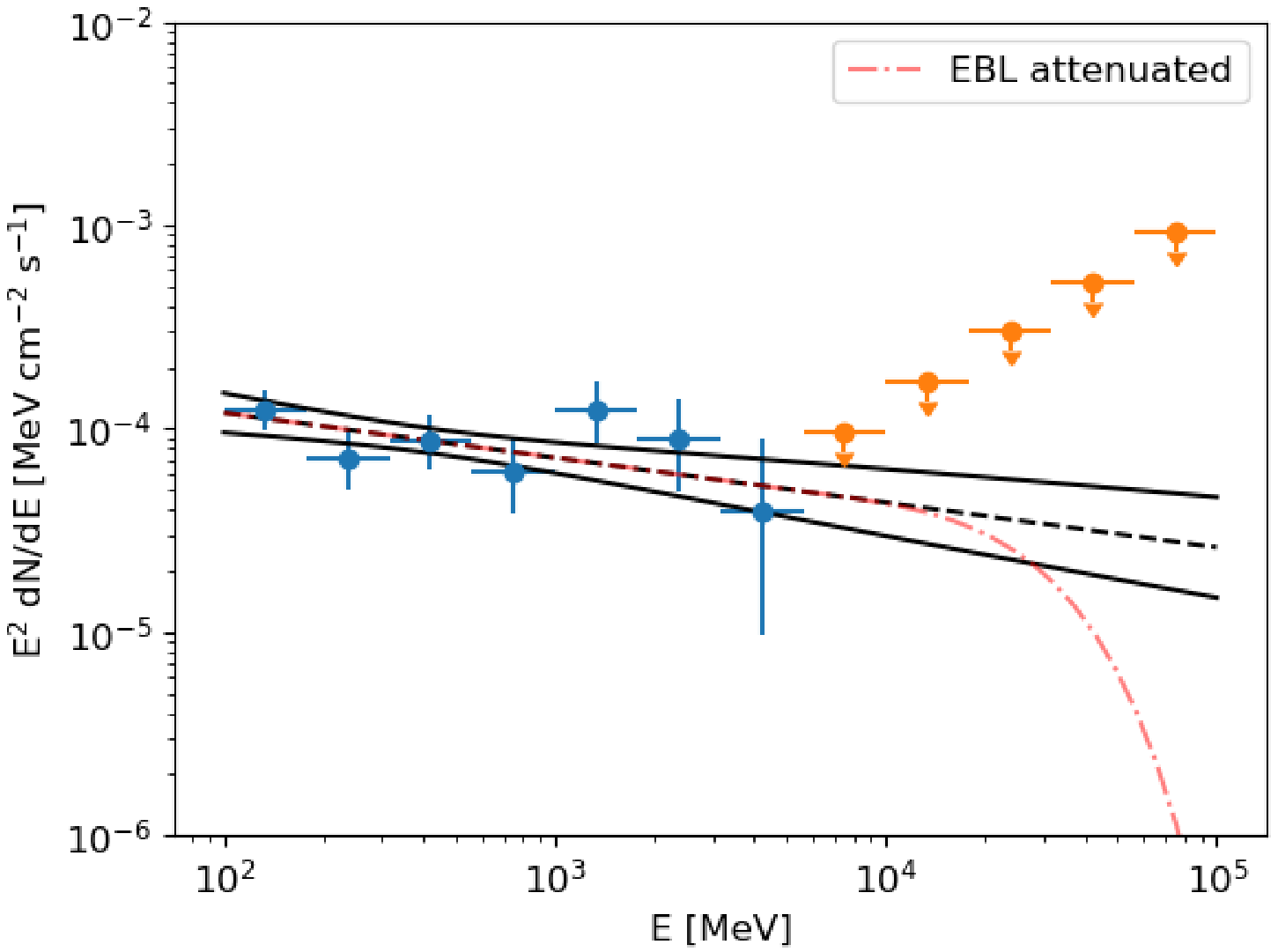}
\includegraphics[width=65 mm, angle= 0]{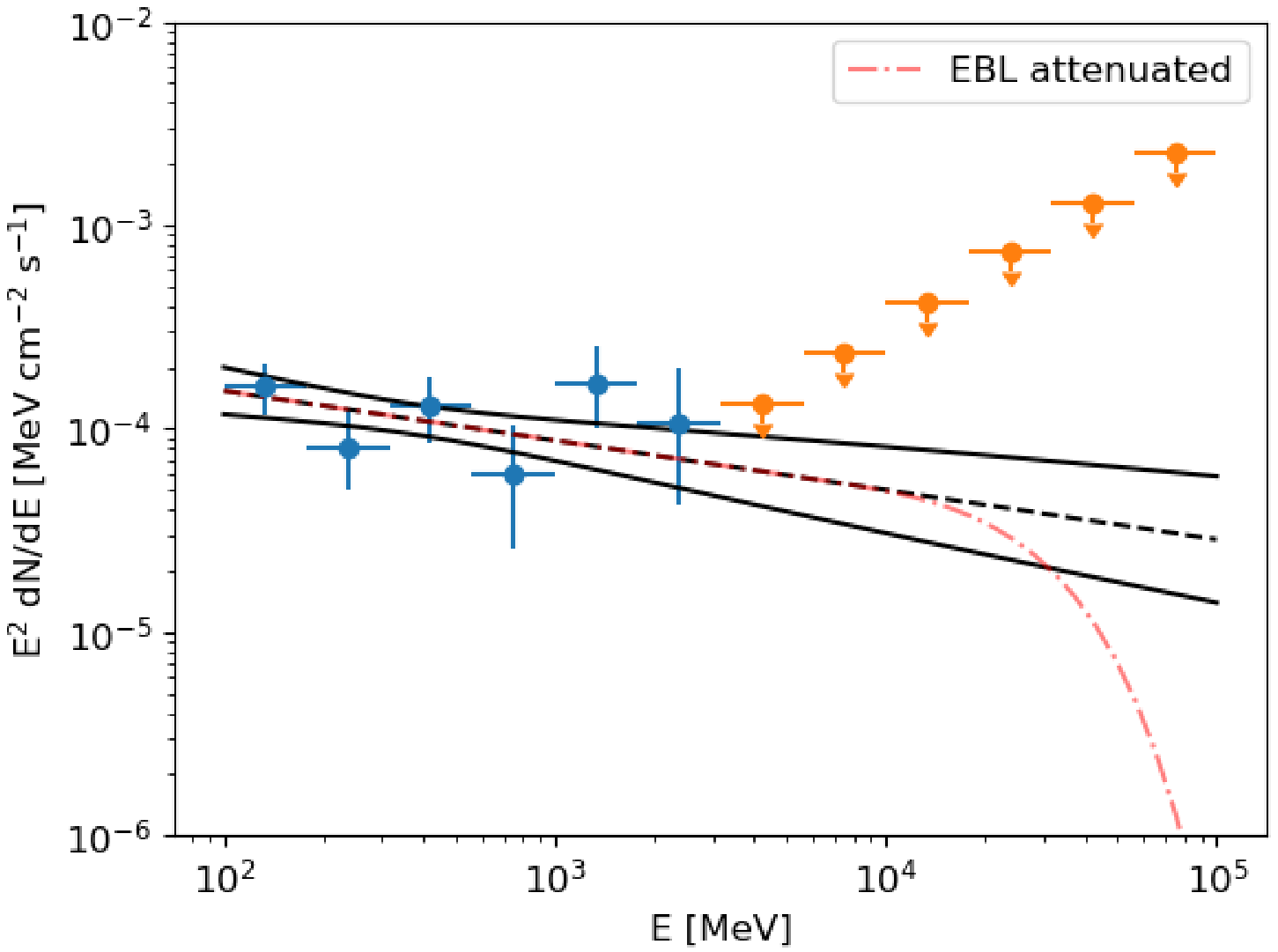}
\includegraphics[width=115 mm, angle= 0]{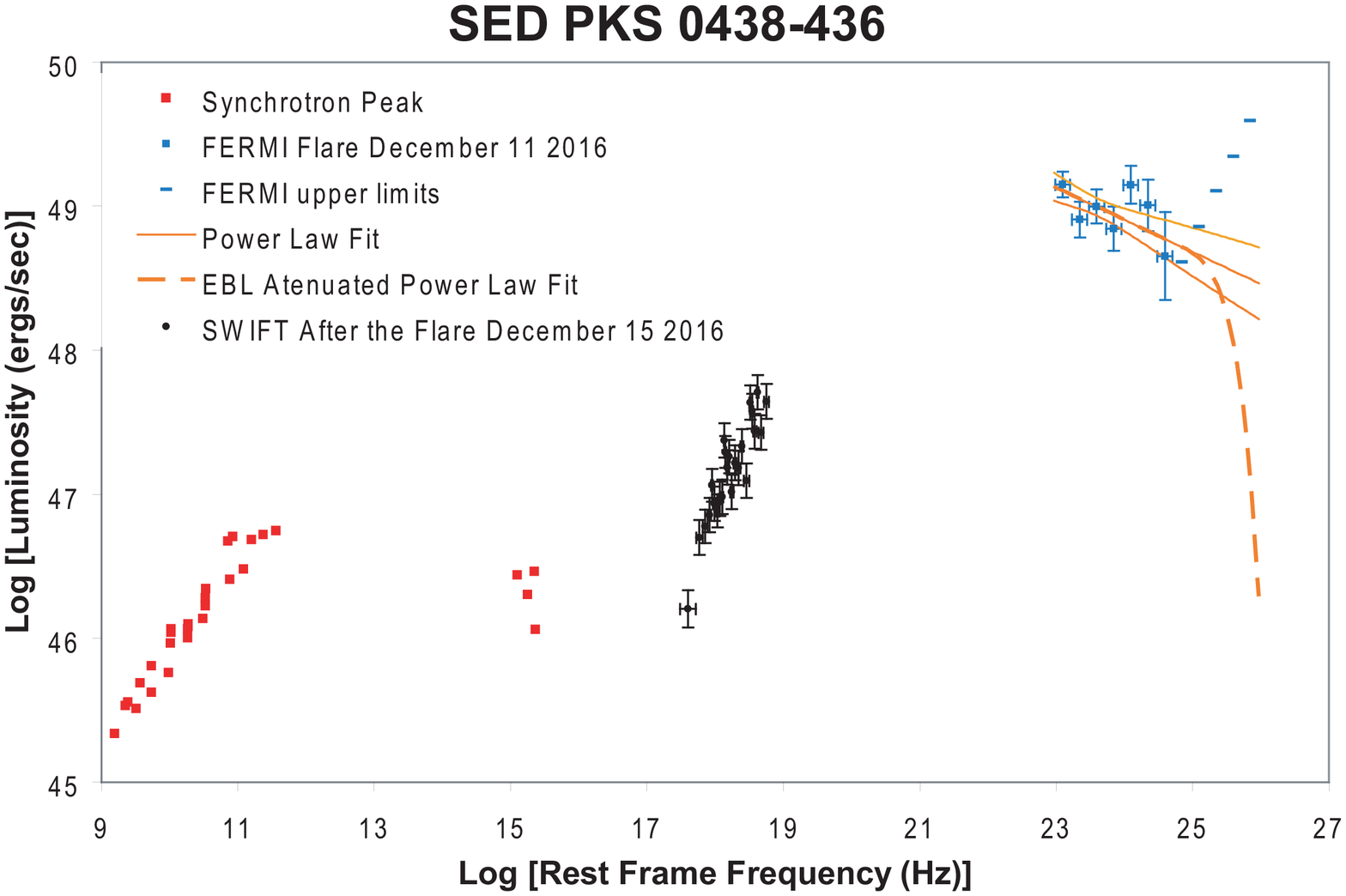}
\caption{\footnotesize{The top frame is the 100 MeV - 100 GeV light
curves in terms of apparent luminosity and count rate. The middle
frames are the spectral fits averaged over the 54 hours of the flare
(left), and the $a$ time range ( right). The red dot-dashed line
corresponds to the best fit model attenuated by the EBL, see the
text for more details. The bottom frame is an SED (not
contemporaneous) from the radio band to $\gamma$-rays.}}
\end{center}
\end{figure}

\begin{table}
    \caption{$Fermi-$LAT Spectral Analysis Summary}
    \tiny
    \centering
    \begin{tabular}{ccccccccc}
        \hline
        Time range  & start  & stop  &  delta T & $\alpha_{\gamma}$& Number & $\sqrt{\rm{TS}}$ & Luminosity  & Truncation Energy   \\
        &          &          &             &        & of  photons   &  statistical            &        &      \\

        &  MJD &  MJD &  hours &       &  detected  & significance & $10^{49}$ ergs~s$^{-1}$ & \\
        \hline
        flaring & 57733.25  & 57735.50 & 54 & $2.2 \pm 0.1$ & $78$ & 10.90  &$4.8\pm1.0$ & 100 GeV\\

        a       &57733.25  & 57734.00  & 18 & $2.2 \pm 0.2$ & $41$ & 8.30  &$5.8\pm1.6 $ & 100 GeV\\\

        b       &57734.00  &57734.75   & 18 & $2.4 \pm 0.4$ & $14$ & 3.38    &$2.2\pm1.1 $& 100 GeV\\\

        c       &57734.75  &57735.50   & 18 & $2.0 \pm 0.2$ & $22$ & 6.65  & $6.6\pm3.0 $ & 100 GeV\\

        flaring & 57733.25  & 57735.50 & 54 & $2.2\pm0.1$ & $77$ & 10.88  & $4.1\pm0.6$& 10 GeV \\

        a       &57733.25  & 57734.00  & 18 & $2.2\pm0.2$ & $41$ & 8.32  & $5.0\pm1.1$ &10 GeV\\\

        b       &57734.00  &57734.75   & 18 &  $2.4\pm0.5$ & $13$ & 3.30    & $2.0\pm0.9$ & 10 GeV\\\

        c       &57734.75  &57735.50   & 18 & $1.9\pm0.3$ & $21$ & 6.57  & $4.9\pm1.6$ & 10 GeV\\
        \hline

    \end{tabular}

\end{table}

We have performed a time resolved spectral analysis for each flaring
bin in the 54-hours time span, and for the full  54-hours time span,
using a {power-law model ( $dN/dE \propto E^{-\alpha_{\gamma}}$),
and freeing all the sources located within a distance of $5
^{\circ}$ from the ROI center. The results are summarized in Table
2. There are 6 sources within the $5 ^{\circ}$ region, and all these
sources show, during the flare period, a flux level at least a
factor of 10 times smaller than the flux of PKS 0438$-$436. The
source is detected with a significance of $\gtrsim 10 \sigma$ during
the full flare period, and with a significance ranging from $\gtrsim
3\sigma$ to $\gtrsim 8\sigma$, during each time bin.
 There is formally no detection above 10 GeV,
only upper limits. The lack of detected photons arises from two
conspiring issues. First is the large distance that makes for low
number statistics even for this very luminous flare. Second is the
EBL (extra background light) opacity. Gamma ray extinction occurs as
a consequence of pair creation in the soft background photon field
\citep{gou67,fra08}. In order to understand the impact of the EBL on
our data, we have calculated the EBL attenuation using the
$\gamma-\gamma$  optical depths  evaluated at z=2.85 from the
template model in \citet{Finke}.\footnote{\url{
http://www.phy.ohiou.edu/~finke/EBL/index.html}} The red dot-dashed
line  in the bottom frames of Figure 3 shows  the expected
attenuation. The EBL starts to be relevant above $\approx$ 70 GeV,
and the upper limits in our SEDs starts at $\approx$ 10 GeV. For
this reason we were unable to disentangle the EBL extinction from
the intrinsic high energy cutoff. The only intrinsic fit warranted
by the data is a power law. Table 2, presents power law fits that
are truncated at 10 GeV and 100 GeV.

\par The SED in Figure 3 are radio data from the NASA
Extragalactic Database, optical from Figure 2, and the FERMI fit to
the 54 hours of December 11 2016 (time range $flaring$ in Table 2).
We also plot the SWIFT spectrum, observed  two days after the flare,
obtained using the Swift XRT products generator
\footnote{\url{http://www.swift.ac.uk/user\_objects/}}
\citep{evans09}. The X-ray spectrum is more luminous and harder than
historical X-ray spectra \citep{cap97,ree00,bro04}. The data are not
contemporaneous, but the strong inverse Compton (IC) $\gamma$-ray
peak $> 5 \times 10^{49}~\rm{ergs~s^{-1}}$ is much stronger than the
synchrotron peak $\sim 2 \times 10^{47}~\rm{ergs~s^{-1}}$,
characteristic of strong quasar gamma ray sources \citep{ghi10}. The
luminosity of the flare in Table 2 is $\sim100$ times the quiescent
upper limit (August 4 2008 - July 4 2009) in \citet{boc16}
indicating extreme variability. Since it is rarely detected, the
time averaged $\gamma$-ray luminosity is not extreme \citep{boc16}.
However, the flare is $\approx 55-65\%$ as luminous (in an 18 hour
window) as the historically large flare of 3C 454.3 even though the
peak of the SED is redshifted out of the FERMI observing window
\citep{abd11}.
\par Considering $\mathcal{D}$, it is of interest to estimate
the intrinsic luminosity of the flare. For an unresolved source
$\mathcal{D}=\delta^{3+\alpha}$, where $\delta$ is the Doppler
factor \citep{lin85}. Based on Table 2 $\alpha \equiv
\alpha_{\gamma}-1 \approx 1.2$. Thus, $\mathcal{D}\approx 15^{4}
\approx 5 \times 10^{4}$, using the average of $\delta=15$ for high
optical polarization quasars estimated from time variability
\citep{hov09}. This implies an intrinsic luminosity of the flare
$\approx 10^{45} \rm{ergs/s}\sim0.01\overline{Q}$. These radiation
losses are easily sustainable by a jet with a power similar to the
long term time average.

\section{Conclusion}
This Letter is the the first comprehensive description of PKS
0438$-$436.. In Section 2, we provided a JVLA image of the radio
source that delineates the kpc structure for the first time. In
Equations (3)-(4) we estimate $\overline{Q} = 1.48 \pm 0.72 \times
10^{47}~\rm{erg~s^{-1}}$ near the maximum value known for quasars
\citep{wil99}. In section 3, we describe the optical spectrum. We
estimate $L_{\mathrm{bol}} $ from the CIV BEL in two different
epochs: $L_{\mathrm{bol}} = 4.7\pm 1.0 \times 10^{46} \rm{ergs/s}:\,
\rm{October\, 2003},\; L_{\mathrm{bol}} = 8.1 \pm 1.7 \times 10^{46}
\rm{ergs/s}:\; \rm{November\, 1977}$. We also estimate
$L_{\mathrm{bol}} $ from the far UV continuum in a low state in
September 2002: $L_{\mathrm{bol}} =4.5 \pm 0.8 \times 10^{46}
\rm{ergs/s}$. There is agreement with the BEL estimate a year later.
This indicates the extremely large normalized jet power,
$\overline{Q}/L_{\mathrm{bol}}= 3.3 \pm 2.6 $, amongst the most
extreme known for quasars \citep{pun07}. In spite of this, PKS
0438$-$436 has only been detected once in the $\gamma$-rays. This is
likely an extreme manifestation of external Compton scattering in a
relativistic jet for which the apparent luminosity is very sensitive
to the geometry \citep{der93}.
\begin{acknowledgements}This work was partially based on observations
collected at the European Organisation for Astronomical Research in
the Southern Hemisphere under ESO programme(s) 69.A-0051(A) and
072.A-0346(A). Based on data obtained from the ESO Science Archive
Facility under request number 295401. This work was supported by the
National Radio Astronomy Observatory, a facility of the National
Science Foundation operated under cooperative agreement by
Associated Universities, Inc., Project 15A-105. Partial
funding for this work was provided by ICRANet.
\end{acknowledgements}

\end{document}